\documentclass[manuscript,screen,nonacm]{acmart}

\renewcommand\footnotetextcopyrightpermission[1]{} 
\usepackage{longtable}
\usepackage{float}
\usepackage{booktabs}
\usepackage{longtable}
\usepackage[normalem]{ulem}
\usepackage{soul}
\renewcommand{\arraystretch}{1.2} 

\AtBeginDocument{%
  }

\setcopyright{acmlicensed}
\copyrightyear{2025}
\acmYear{2025}
\acmDOI{XXXXXXX.XXXXXXX}

\acmConference[CHI '25]{ACM CHI Conference on Human Factors in Computing Systems}{April 26 – May 1, 2025,
  2025}{Yokohama, Japan}
\acmISBN{978-1-4503-XXXX-X/18/06}

\begin{document}

\title{Improving governance outcomes through AI documentation: Bridging theory and practice}

\author{Amy A. Winecoff}
\email{awinecoff@cdt.org}
\orcid{0000-0002-2594-4126}
\affiliation{%
  \institution{Center for Democracy \& Technology}
  \city{Washington} 
  \state{DC}
  \country{USA}
}

\author{Miranda Bogen}
\email{mbogen@cdt.org}
\affiliation{%
  \institution{Center for Democracy \& Technology}
  \city{Washington}
  \state{DC}
  \country{USA}
}

\renewcommand{\shortauthors}{Winecoff \& Bogen}

\begin{abstract}
Documentation plays a crucial role in both external accountability and internal governance of AI systems. Although there are many proposals for documenting AI data, models, systems, and methods, the ways these practices enhance governance as well as the challenges practitioners and organizations face with documentation remain underexplored. In this paper, we analyze 37 proposed documentation frameworks and 22 empirical studies evaluating their use. We identify several pathways or "theories of change" through which documentation can enhance governance, including informing stakeholders about AI risks and applications, facilitating collaboration, encouraging ethical deliberation, and supporting best practices. However, empirical findings reveal significant challenges for practitioners, such as insufficient incentives and resources, structural and organizational communication barriers, interpersonal and organizational constraints to ethical action, and poor integration with existing workflows. These challenges often hinder the realization of the possible benefits of documentation. We also highlight key considerations for organizations when designing documentation, such as determining the appropriate level of detail and balancing automation in the process. We conclude by discussing how future research can expand on our findings such as by exploring documentation approaches that support governance of general-purpose models and how multiple transparency and documentation methods can collectively improve governance outcomes. 

\end{abstract}



\received{TBD}
\received[revised]{TBD}
\received[accepted]{TBD}

\maketitle

\section{Introduction}

Transparency in AI is widely regarded as essential for calibrating trust and supporting accountability \cite{liao2023ai}. By providing stakeholders with a clear view of an AI system's composition, operation, and development process, transparency allows for informed oversight and critical evaluation. When AI systems are thoroughly documented, these documentation artifacts offer invaluable insights into important system features such as training data, algorithms, and risk management strategies. This information can support several groups of external stakeholders. Documentation can help downstream deployers and users understand how the system functions and what risks it conveys. It can also assist policymakers and researchers in holding companies accountable for the negative consequences of their AI technologies.

Yet the importance of documentation extends beyond external accountability.\footnote{Though the concepts of documentation and transparency are often blurred, complete documentation about an AI system does not guarantee that information will be made available to interested stakeholders, nor do transparency artifacts always reflect the full or most relevant details of an AI system even though such details may have been captured in internal documents. We distinguish between the concepts in this report in order to provide greater clarity about the presumptive goals of documentation and an exploration of what practices may more likely support those objectives.} Internally, documentation serves as a critical tool for managing AI systems throughout their lifecycle for a wide range of stakeholders. For example, documentation could help downstream technical practitioners understand the strengths, limitations, and risks of training a model on a given dataset. It might also help compliance and governance teams assess a potential system use case's compliance with company policies or legal requirements. Or the process of producing documentation might lead practitioners to make different decisions about how to sufficiently mitigate potential harms. Robust documentation can improve the efficiency and quality of other important AI governance practices and processes, such as internal and third-party auditing, harm identification, risk triaging, and compliance functions, and can help practitioners better implement risk management practices, such as system evaluation and risk mitigation approaches. In this way, documentation serves as a foundational building block of a multifaceted AI governance approach, both providing important inputs to other governance processes and creating a scaffold on which additional governance practices can build over time.

Despite its clear benefits, creating effective documentation is a complex task. No single form of documentation can meet the diverse needs of all stakeholder groups equally. For example, while detailed technical reports like those for Llama-2\cite{touvron2023llama2} and GPT-4\cite{openai2023gpt4v}, which can exceed 50 pages and contain dense technical language, are valuable for AI researchers, they may not be accessible to non-technical stakeholders. Similarly, documentation intended to satisfy compliance requirements might fall short in providing practical guidance for those deploying AI systems.

To maximize the impact of documentation on AI governance, organizations must carefully define their goals and identify who will produce, maintain, and use the documentation. Tailoring the scope, level of detail, and format of documentation to suit the intended audiences and purposes is crucial. Effective documentation design requires balancing the needs of both the documentation process and the resulting artifacts, ensuring that they collectively support the organization's governance objectives.

In this paper, we explore how documentation processes and artifacts can serve as one critical building block in supporting AI governance and risk management goals.  Our research draws from a qualitative analysis of academic and gray literature,\footnote{Gray literature refers to publications that have not been peer-reviewed, but that present detailed theories or research findings} including 37 proposed documentation frameworks and 22 papers with relevant empirical findings. From this analysis, we identify four "theories of change" or pathways through which documentation could improve AI governance, including informing stakeholders about responsible use, facilitating collaboration on AI risks, encouraging ethical deliberation, and improving overall governance and development practices. Our analysis also identified several challenges that might prevent documentation from effectively supporting governance aims. These include insufficient incentivization and resourcing for documentation, structural and organizational barriers to effective collaboration, interpersonal and organizational constraints to ethical action, and poor integration with practitioners' day-to-day workflows and processes.

We also explored considerations or trade-offs organizations face when designing and implementing documentation practices to balance competing priorities or documentation goals. These considerations include balancing customization with standardization, deciding between single or multiple forms of documentation to address different audiences, determining the extent of detail to include, choosing the appropriate level of automation to employ in the documentation process, and selecting between static or interactive formats.

The primary contribution of our research is that our synthesis identifies patterns, gaps, and emerging trends across individual publications about documentation approaches and evaluations. By offering a holistic analysis of how AI documentation can support robust governance in real-world contexts, we clarify which documentation practices are likely to be most effective, under what conditions, and for which stakeholders. As such, our research can serves as a valuable resource for both researchers and practitioners seeking to refine and enhance their approaches to AI documentation in applied AI development contexts. We conclude with a discussion of how future research can build on or expand upon the themes we explore throughout the paper.

\section{What organizations could document about AI systems}
Documenting the various components of AI systems—data, models, systems, and methods—is essential for effective governance and informed decision-making. This documentation not only provides insight into the development process but also helps internal and external stakeholders understand the characteristics and potential risks associated with AI systems.

Data is a fundamental component of AI systems since it significantly influences model behavior and performance. When models are trained on datasets that don't align with their deployment contexts or that contain biases, the outcomes can be problematic, reinforcing those biases or leading to poor performance \cite{gebru2021datasheets}. Data documentation, which might include details about dataset characteristics, collection methods, preprocessing, and known limitations, allows practitioners to identify potential issues early on, supporting better model performance and informing the decisions of those who use the documentation \cite{bender2018data, chmielinski2020dataset, diaz2023sound, gebru2021datasheets, marone2023data, miceli2021documenting, papakyriakopoulos2023augmented, pushkarna2022data, roman2023open, rostamzadeh2022healthsheet, soh2021building, stoyanovich2019nutritional, subramaniam2023comprehensive, sun2019mithralabel, zheng2022network}. By tailoring documentation to the specific data type and application context, organizations can enhance its utility, drawing on established data documentation frameworks to guide this process.

Model documentation also enhances AI development and governance, providing a window into a model's creation, performance, and risks \cite{crisan2022interactive, mcmillan2021reusable, mitchell2019model, shen2022model, stoyanovich2019nutritional}. Documenting aspects such as the model's architecture, training procedures, intended use cases, and evaluation methods enables practitioners to assess potential risks and responsibly adapt models to specific contexts. For models developed in stages—such as those pretrained on large datasets and later fine-tuned—documenting each stage is essential \cite{gilbert2023reward}. This helps downstream users understand the development pipeline and its implications for their intended use.

AI services or applications in the real world typically consist of complex systems rather than isolated models. These systems often include multiple components, such as machine learning models, trust and safety filters, third-party software integrations, and user interfaces. The interaction among these components significantly affects the system's overall performance and risk profile — especially because the properties or behavior of individual components may change when integrated into a system \cite{smart2024sociotechnical}. Recognizing that AI systems are more than just the sum of their parts, researchers have suggested that organizations should document not only each individual component but also the system as a whole \cite{arnold2019factsheets, baracaldo2022towards, brajovic2023model, procope2022system, yang2018nutritional, blasch2021multisource}. System documentation could provide an overview, including objectives, inputs, outputs, and a diagram illustrating how the components interact \cite{procope2022system}. This comprehensive approach ensures that practitioners have a clear understanding of the system's behavior in its entirety.

Organizations may also find it important to document the processes and procedures that shape the development and deployment of AI systems and their components. This documentation could include how humans annotated or evaluated training or evaluation data \cite{shimorina2021human}, any privacy, legal, and ethical reviews or audits conducted \cite{raji2020about}, and any risk mitigations that were applied.
Investing time in this type of documentation may be especially beneficial because it cannot be deduced merely from examining the system components. If such meta-data about the process is not documented throughout the development lifecycle, it may be very difficult to reconstruct later \cite{belz2023missing, reid2023right}. As a result, it is ideal to produce documentation on development processes concurrently with the development of the system.

Organizations may also benefit from documenting machine learning methods \cite{adkins2022prescriptive, baracaldo2022towards, sokol2020explainability, tagliabue2021dag}, such as techniques applied in computer vision applications \cite{adkins2022prescriptive} and strategies for enhancing model explainability \cite{sokol2020explainability}. Documentation of the application of the AI model or system to specific AI tasks or use cases \cite{hupont2022documenting, mohammad2022ethics} can also help practitioners understand the benefits and limitations of these applications. Information about methods and processes can provide valuable insights into what potential approaches a team might take to develop an AI system. Such insights can help practitioners determine when particular approaches are more or less appropriate and what risks different approaches may confer.

Within companies, many stakeholders may contribute to and use documentation \cite{micheli2023landscape}. Those working on data collection, annotation, and curation might contribute to data documentation, while engineers and data scientists might document and use information related to data, models, and application infrastructure. UX designers, researchers, and product managers might rely on this documentation to inform product designs and create clear public explanations. Professionals focused on responsible AI, governance, and compliance could use documentation to identify and manage risks. Organizational leaders, in turn, may use these insights to make critical decisions about resource allocation, risk management, and the launch of AI-powered products.

We can categorize documentation stakeholders into two broad groups: \textbf{documentation producers} and \textbf{documentation consumers}. Documentation producers are those who actively contribute to the generation of documentation artifacts, while documentation consumers are those who read and use it. This grouping simplifies some important distinctions.\footnote{Practitioners sometimes create documentation for their own use, while in other instances, they produce one type of documentation for others and rely on a different type created by someone else. For example, a compliance team might use documentation from a technical team to understand how a system was developed and functions in order to assess compliance concerns. At the same time, the compliance team might generate documentation for technical teams, outlining the limitations that apply to the use of a system or component. In this way, the same stakeholder could be both a consumer and a producer of documentation, even for the same system or component.} Still, distinguishing between consumers and producers serves as a helpful heuristic for understanding the opportunities and tensions between groups with both shared and distinct goals. We consider both producers and consumers and their interplay in our analysis.

We also note that "AI documentation" could refer to either the process of producing documentation or the resulting artifacts. While documentation artifacts can inform stakeholders about the responsible use of AI, the process of creating these documents can also institutionalize best practices and foster a culture of risk management. Therefore, when considering how documentation can improve governance outcomes, we address both artifacts and processes.

\section{Methods}
\subsection{Publication Sampling}

We employed a purposeful sampling approach \cite{palinkas2015purposeful, suri2011purposeful} to identify publications. This method was particularly well-suited to our study because it allowed us to target information-rich frameworks and empirical studies clearly aligned with our research goals of understanding the functions documentation plays in supporting governance and the challenges associated with using documentation as a foundational governance tool. By prioritizing publications that offered insights into documentation frameworks, evaluation methods, and practical applications, we ensured extensive coverage of existing theories and practices while maintaining direct relevance to our research questions. Furthermore, our iterative sampling process enabled us to refine our dataset as our analysis progressed, incorporating studies that contributed meaningful data or theoretical perspectives. Our approach contrasts with systematic sampling, which establishes strict inclusion and exclusion criteria beforehand. While systematic sampling might have offered more exhaustive coverage, it likely would have included less relevant papers that did not align as closely with our objectives and would not have allowed us to expand our sample in response to emerging findings.

We opted to include both proposed frameworks and empirical evaluations in our analysis rather than limiting our focus to empirical studies. This decision stemmed from the recognition that the goals of a documentation framework and its real-world implementation can often diverge. For instance, a framework that generates high-quality documentation in a research context may fall short when applied in practical settings. By examining both the intended goals of proposed frameworks and the empirical evidence, we aimed to better understand and highlight the gap between these two dimensions, informing future designs and evaluations. 

We identified an initial seed sample of publications from a systematic review focused on AI documentation methods relevant to EU regulation \cite{micheli2023landscape}. We cross-referenced this initial sample against the references used by the Partnership on AI in developing their ABOUT ML framework,\footnote{https://partnershiponai.org/workstream/about-ml/} which seeks to provide technology organizations with guidance on what to document about their AI systems. We chose these initial sources because they allowed us to focus on publications that aimed to establish best practices for industry documentation and address policy initiatives geared toward institutionalizing these practices. We reviewed the references of these publications and consulted with subject matter experts in industry and academia with experience producing or using documentation to identify additional publications.

This process yielded a sample of 37 proposed frameworks, eight of which included empirical evidence related to their implementation. To expand our sample of empirical findings, we again consulted with subject matter experts and reviewed the citations of the frameworks for references to further empirical work. We also conducted searches in the archives of the ACM Conference on Human Factors in Computing Systems and the ACM Conference on Computer Supported Cooperative Work using terms such as "datasheet," "model card," "AI documentation," "model documentation," "data card," and "data documentation." Together, these methods allowed us to identify 14 additional publications with empirical findings. Our sampling concluded upon reaching theoretical saturation, where further sampling no longer yielded new themes, supporting the adequacy and richness of our chosen publications.

Recognizing the importance of applied perspectives, we included non-peer-reviewed papers, such as technical whitepapers published by industry actors, acknowledging that they often provide insights not found in academic literature. To maintain a targeted focus on publications examining the governance role of documentation rather than the technical aspects of specific tools, we excluded papers discussing enabling technologies (e.g., automated documentation libraries or supporting software) or those addressing broader topics like fairness checklists not specific to AI documentation.

\subsection{Data Analysis}
We employed an abductive approach for our data analysis \cite{tavory2014abductive, timmermans2012theory}, a qualitative method that involves alternating between theoretical concepts identified deductively and codes generated inductively through data review, similar to grounded theory methods. \cite{charmaz2014constructing, strauss1990basics, glaser1978theoretical, glaser1968discovery}. Guided by this approach, we initially applied codes aligned with our objectives of understanding how documentation can support robust AI governance and identifying barriers to its effectiveness. At the same time, we remained open to adjusting our coding focus based on insights gained through inductive analysis of our publication sample. Our analysis was refined iteratively through ongoing discussions and repeated reviews of the sample publications in alignment with these objectives.

In the initial coding phase, the first author applied descriptive codes to papers that aligned with our research objectives. These codes included terms like “discoverability,” “standardization,” “intended use,” “incentivization,” “documentation errors." During the review of these descriptive codes, both authors observed emerging themes related to both theories of change and challenges. For example, we noted that both the codes "identify intended purpose" and "specify out of scope uses" both pertained to informing downstream documentation consumers. Following this discussion, we organized our descriptive codes into four theory of change categories: informing downstream stakeholders, facilitating collaboration and communication, promoting ethical deliberation, and supporting governance and development best practices. We grouped codes related to challenges into four overarching categories: insufficient incentives and resources, structural and organizational obstacles to effective communication, limited ethical awareness and barriers to ethical action, and inadequate integration with existing workflows and organizational practices. We note that in some cases, challenges are more relevant to specific theories of change, and in other cases, challenges are more generally applicable. We discuss these points of connection in the results.

The inductive aspects of our analysis highlighted that authors of documentation publications made varied design choices aligned with the goals of their proposed frameworks. Consequently, we expanded our focus to examine the design tensions and considerations that surfaced within and across these papers. To capture these dynamics, we grouped descriptive codes related to design choices, particularly those that appeared to exist in tension with one another. These considerations included the balance between customization and standardization in addressing specific components or use cases, the extent of audience-specific tailoring in documentation artifacts, the level of interactivity provided, the amount of detail included, and the degree of automation in the documentation process.

\subsection{Positionality}
The authors of this study work at a civil society organization specializing in technology policy. Previously, the first author developed machine learning models, and the second author focused on AI governance and policy within technology companies. Both authors have experience conducting research in human-centered and responsible AI and have advised government and industry on implementing risk management practices for AI systems. Our academic backgrounds in psychology and public policy inform our understanding of AI as a sociotechnical system and AI governance as a sociotechnical process. These perspectives inevitably shaped our selection of publications, the themes we chose to explore, and our synthesis of the results. Given that our analysis approach is interpretive, researchers with different perspectives and backgrounds may have arrived at different conclusions when analyzing the same dataset or found different publications more meaningful to the overall research aims.

\section{Results}
\subsection{Publication Sample}
The publications included in our analysis are listed in Table \ref{tab:ai_docs}. For each framework, we classified whether the approach primarily focused on data; models; systems; or methods, tasks, and processes. When a framework did not fit neatly into one of these categories, we assigned it to the most appropriate category or categories based on its primary focus.

We also categorized the type of evaluation each framework or empirical research study employed. Frameworks employing a "feasibility analysis" are those where the framework's authors or another group applied the framework to create documentation for a hypothetical or actual dataset, model, system, or method. This type of analysis demonstrates that the framework could theoretically be used for its intended purpose but does not involve empirical evaluation with practitioners in research or real-world settings. If a study developed a documentation artifact for the purpose of an empirical study, we classified this as part of the empirical study rather than as a feasibility analysis, as empirical studies offer a more rigorous evaluation.

We classify publications as employing a "practitioner lab study" when the evaluation involved practitioners within a controlled research setting. We classify "practitioner real-world studies" as those examining practitioner methods and practices within their real-world work environments. Both lab and real-world studies have unique strengths, and neither is inherently more rigorous or useful than the other. We define "artifact studies" as studies of publicly available documentation artifacts such as Github repository documentation or Hugging Face model cards. 

In some instances, framework authors mentioned consulting relevant stakeholders during the design or refinement of their framework. However, if these consultations were only briefly mentioned, we do not classify the work as including a practitioner study.

\renewcommand{\arraystretch}{1.2} 

\begin{longtable}{c l c l l}
\caption{List of Publications} \\
\toprule
\textbf{Num.} & \textbf{Author} & \textbf{Year} & \textbf{Framework Type} & \textbf{Evaluation Type} \\
\midrule
\endfirsthead

\toprule
\textbf{Num.} & \textbf{Author} & \textbf{Year} & \textbf{Framework Type} & \textbf{Evaluation Type} \\
\midrule
\endhead

\midrule \endfoot
\bottomrule \endlastfoot

1  & Adkins et al.           & 2022 & Method, process, or task  & Feasibility analysis \\
2  & Ahlawat, Winecoff, \& Mayer & 2024 & Evaluation only  & Practitioner real-world study \\
3  & Arnold et al.           & 2019 & System                    & Feasibility analysis \\
4  & Baracaldo et al.        & 2022 & System                    & Feasibility analysis \\
5  & Bender \& Friedman      & 2018 & Data                      & Feasibility analysis \\
6  & Bhat et al.             & 2023 & Evaluation only           & Practitioner lab study \\
7  & Blasch et al.           & 2020 & System                    & Feasibility analysis \\
8  & Boyd                   & 2021 & Evaluation only           & Practitioner lab study \\
9  & Brajovic et al.         & 2023 & System                    & Feasibility analysis \\
10  & Chang \& Custis         & 2022 & Evaluation only           & Practitioner real-world study \\
11 & Chmielinski et al.      & 2020 & Data                      & Feasibility analysis \\
12 & Chmielinski et al.      & 2024 & Method, process, or task  & None \\
13 & Crisan et al.           & 2022 & Model                     & Practitioner lab study \\
14 & Díaz et al.             & 2023 & Data                      & Feasibility analysis \\
15 & Gebru et al.            & 2021 & Data                      & Feasibility analysis \\
16 & Geiger et al.           & 2020 & Evaluation only           & Artifact study \\
17 & Gilbert et al.          & 2023 & System                    & Feasibility analysis \\
18 & Heger et al.            & 2022 & Evaluation only           & Practitioner real-world study \\
19 & Hind et al.             & 2019 & Evaluation only           & Practitioner real-world study \\
20 & Holland et al.          & 2018 & Data                      & Feasibility analysis \\
21 & Hupont \& Gomez         & 2022 & Method, process, or task  & Feasibility analysis \\
22 & Liang et al.            & 2024 & Evaluation only           & Artifact study \\
23 & Liao et al.             & 2023 & Evaluation only           & Practitioner lab study \\
24 & Marone \& Van Durme                & 2023 & Data                      & Feasibility analysis \\
25 & McMillan-Major et al.          & 2021 & Data, Model               & Feasibility analysis \\
26 & McMillan-Major, Bender, \& Friedman          & 2024 & Data                      & Practitioner lab study \\
27 & Miceli et al.           & 2021 & Evaluation only           & Practitioner real-world study \\
28 & Mitchell et al.         & 2019 & Model                     & Feasibility analysis \\
29 & Mohammad                & 2022 & Method, process, or task  & Feasibility analysis \\
30 & Moore, LIao, \& Subramonyam & 2023 & Evaluation only           & Practitioner lab study \\
31 & Nunes et al.            & 2022 & Evaluation only           & Practitioner lab study \\
32 & Papakyriakopoulos et al. & 2023 & Data                      & Feasibility analysis \\
33 & Pepe et al.             & 2024 & Evaluation only           & Artifact study \\
34 & Procope et al.          & 2022 & System                    & Feasibility analysis \\
35 & Pushkarna et al.        & 2022 & Data                      & Practitioner real-world study \\
36 & Raji \& Yang            & 2020 & System                    & None \\
37 & Reid \& Williams        & 2023 & Evaluation only           & Practitioner real-world study, Artifact study \\
38 & Richards et al.         & 2020 & System                    & Feasibility analysis \\
39 & Roman et al.            & 2023 & Data                      & Feasibility analysis \\
40 & Rostamzadeh et al.      & 2022 & Data                      & Practitioner lab study \\
41 & Shen et al.             & 2022 & Model                     & Practitioner lab study \\
42 & Shimorina \& Belz       & 2021 & Method, process, or task  & None \\
43 & Soh                    & 2021 & Data                      & None \\
44 & Sokol \& Flach          & 2020 & Method, process, or task  & Feasibility analysis \\
45 & Srinivasan et al.       & 2021 & Data                      & Feasibility analysis \\
46 & Stoyanovich \& Howe     & 2019 & Data, Model               & Feasibility analysis \\
47 & Subramaniam et al.      & 2023 & Data                      & Practitioner lab study \\
48 & Sun et al.              & 2019 & Data                      & Feasibility analysis \\
49 & Tagliabue et al.        & 2021 & Method, process, or task  & Feasibility analysis \\
50 & Yang et al.             & 2018 & Method, process, or task  & Feasibility analysis \\
51 & Zheng et al.,                  & 2022 & Data                      & Practitioner lab study \\
\label{tab:ai_docs}
\end{longtable}

\subsection{Theories of change}
While empirical evidence directly linking documentation artifacts or processes to improved governance outcomes is limited, proposed documentation frameworks offer compelling theories of change about how documentation can positively a multifaceted impact governance approach. Our analysis revealed four such theories of change: informing downstream users about system development and associated risks, encouraging ethical reflection among practitioners, facilitating communication among stakeholders, and enhancing AI development and governance overall.



\subsubsection{Theory of change \#1: Documentation could inform downstream stakeholders about effective and responsible use.}

Documentation can provide essential background context on the system or component's development, informing appropriate integration of data, models, methods, or systems into technologies. For example, practitioners typically develop systems with specific use cases in mind, and other development teams need to understand these intended use cases to assess alignment with their goals. If aligned, teams can then modify the systems to meet their current objectives. Practitioners argue documentation can also clarify underlying assumptions, such as data collection methods and representativeness, which are critical for assessing the suitability of a dataset for specific purposes \cite{reid2023right, rostamzadeh2022healthsheet, zheng2022network}. Also, documentation on background context can help identify potential issues prior to deployment, such as mismatches between the original training data and downstream practitioners' intended deployment setting \cite{boyd2021datasheets}. 

Moreover, documentation could provide information that helps downstream practitioners make sound implementation decisions. For example, documentation could help practitioners find and compare system components aligned with their goals, including alternatives to AI approaches \cite{heger2022understanding, reid2023right, shen2022model, zheng2022network}. It might also provide statistical summaries of data \cite{holland2018dataset} or or model performance evaluations such as area under the curve (AUC) \cite{mitchell2019model} to help inform developers’ choices of how to train models or implement appropriate guardrails. 

Documentation could also help stakeholders ensure that planned use cases comply with legal constraints, regulatory requirements, and organizational policies. For example, certain datasets may be restricted from use if the data were collected without affirmative consent or contain personally identifiable or copyrighted information. Without this awareness, practitioners might inadvertently train a model on such data, only to later discover that they cannot use the model due to policy violations \cite{roman2023open}. Practitioners see avoiding disciplinary action for non-compliance with company policies as a motivating benefit of documentation \cite{heger2022understanding}.

The type of information that is most useful to downstream stakeholders may depend on whether the documentation pertains to data, models, systems, or methods and tasks. For example, system documentation can help downstream stakeholders understand functionality and risks that emerge from the interactions of underlying components \cite{arnold2019factsheets, procope2022system}. Data documentation, on the other hand, can help downstream practitioners assess the downstream impact of upstream inputs, such as potentially privacy-compromising data inputs or sampling methods that could result in unfair user treatment \cite{gebru2021datasheets, boyd2021datasheets}.

\subsubsection{Theory of change \#2: Documentation could support cross-functional collaboration and communication about AI risks.}
Since no single group oversees all stages of development, effective documentation can facilitate both indirect and direct communication among these diverse teams. Documentation artifacts can convey essential information across organizational boundaries, helping stakeholders who may not interact regularly to align their understanding of the system through documentation artifacts. Documentation can also prompt and support conversations among stakeholders directly, enabling them to work towards common goals despite their diverse backgrounds, frames of reference, or areas of expertise. This bridging function of documentation could break down organizational silos \cite{chmielinski2024clear}  and promote effective collaboration across different teams \cite{crisan2022interactive, gilbert2023reward, heger2022understanding, mohammad2022ethics, pushkarna2022data, raji2020about, richards2020methodology, shen2022model, srinivasan2021artsheets}.

At a basic level, AI documentation could serve as an initial point of reference for downstream practitioners, offering an overview of the system that allows them to build foundational knowledge before engaging with developers for more in-depth discussions \cite{liao2023designerly, heger2022understanding}.  Rather than acting as a comprehensive, self-contained repository of information, documentation artifacts could provide an initial layer of understanding, enabling downstream practitioners to identify relevant areas for further exploration and formulate more informed questions when collaborating with developers. This could foster more productive and targeted interactions between teams, enhancing their ability to work with or govern AI systems efficiently. Empirical research suggests that practitioners often approach documentation this way. For instance, UX professionals often prefer direct discussions with data scientists rather than relying solely on documentation artifacts \cite{liao2023designerly}. Similarly, many AI practitioners review existing documentation only briefly before seeking further clarification through meetings or discussions \cite{heger2022understanding}. 

In more robust forms, documentation can facilitate deeper deliberation among stakeholders about the system's design and the documentation process itself \cite{pushkarna2022data, shen2022model}. For example, documentation has been shown to aid teams in collectively reflecting on socially constructed concepts, such as gender, and making informed decisions about how to annotate such attributes in datasets \cite{pushkarna2022data}. Also, the process of creating documentation can help stakeholders identify and discuss trade-offs between competing objectives in system design \cite{shen2022model}.

By having some stakeholders define requirements for documentation artifacts, others generate content, and still others assess the quality of the artifact, the documentation process can help stakeholders understand each others’ needs and constraints in ways that could promote quality, efficiency, and responsibility in the development process \cite{chang2022understanding}. Over time, engaging with documentation could also improve the technical literacy of stakeholders, enabling them to create products that better meet user needs while minimizing harm \cite{liao2023designerly}.

\subsubsection{Theory of change \#3: Documentation could prompt practitioners to deliberate and act on the ethical impacts of AI}
As with other tools that surface AI harms \cite{balayn2023fairness}, documentation could encourage practitioners to consider the ethical implications of their systems.

One theory is that when downstream documentation \textit{consumers} consult documentation on data, models, and systems, it can prompt consideration of potential harms. This consideration could lead practitioners to make more careful decisions about if and when to use certain systems or components. For example, one study found that practitioners with access to data documentation were more likely to identify ethical concerns unprompted. However, those without documentation often recognized similar concerns when prompted by the researcher. Despite this, practitioners both with and without access to documentation struggled to develop concrete action plans \cite{boyd2021datasheets}. These findings indicate that while documentation can facilitate ethical deliberation, it is not always essential for ethical reflection and does not consistently lead to actionable solutions.


Another theory is that documentation also might sensitize \textit{producers} to the ethical impacts of their work because documentation frameworks often ask documentation producers to consider and describe the ethical dimensions of their systems. For example, the one framework prompts practitioners to think about privacy implications by inquiring about sensitive information in data \cite{gebru2021datasheets}. Another asks about known biases, ethical issues, and safety concerns \cite{arnold2019factsheets}, and several frameworks encourage reflection on how systems might negatively impact marginalized users or populations \cite{brajovic2023model, mcmillan2021reusable, shen2022model}. 

Some researchers have suggested that documentation might be effective at promoting ethical action, even if it only engages practitioners in critical reflection relatively superficially \cite{bender2018data}. Yet others have taken a more explicit, deliberative approach. Drawing from value-sensitive design \cite{friedman2006value}–a method that helps technologists identify and understand normative judgments in the development process–Shen and colleagues developed the Model Authoring Toolkit to help practitioners consider diverse stakeholder values and deliberate on system design trade-offs \cite{shen2022model}. Their qualitative and survey study of Wikipedia communities found that framing ethical reflection and documentation as a participatory process led to more informed decisions about AI system design and deployment.

\subsubsection{Theory of change \#4: Documentation could catalyze overall improvements in governance and development.}
In applied settings, AI documentation rarely functions in isolation; it often relies on and reinforces complementary practices that could contribute to improved AI outcomes \cite{ahlawat2024minimum}. Our analysis revealed that several ways documentation could improve the rigor, efficiency, or reliability of development and governance practices overall. 

As one example, framework authors posit that documentation could improve the rigor with which practitioners develop AI systems \cite{holland2018dataset}. By requiring practitioners to justify their development choices, documentation may lead to more careful decision-making \cite{bender2018data, chmielinski2024clear, holland2018dataset}. Also, clear and comprehensive documentation supports reproducibility, aiding other practitioners in retraining and validating systems consistently \cite{adkins2022prescriptive, baracaldo2022towards, hind2019experiences}. Since AI practitioners are often driven by a commitment to scientific rigor \cite{winecoff2022artificial} and product quality assurance \cite{ahlawat2024minimum}, documentation that emphasizes these aspects could potentially serve as a "value lever \cite{shilton2013values}," further encouraging engagement with the documentation process.

Documentation might also improve development and governance efficiency. In large organizations, datasets, models, and systems developed for one purpose may also be useful to other teams with similar goals. Documentation can make system components more discoverable, helping practitioners avoid duplicating efforts \cite{heger2022understanding}. Documentation can quickly provide information on the limitations of using a dataset, model, or system \cite{rostamzadeh2022healthsheet}, which could allow practitioners to allocate more time to identifying and addressing any additional risks that may arise when integrating components into new systems. It can also include details relevant to company or legal policies that apply to system components, preventing practitioners from wasting time on products and features that their organizations ultimately won't approve for deployment \cite{roman2023open}. Practitioners also point out that by documenting unsuccessful and successful approaches, organizations can prevent repeated mistakes and identify practices that would be beneficial for the organization to disseminate broadly \cite{chang2022understanding, miceli2021documenting}.

Documentation could also facilitate proactive risk mitigation, which is often more efficient and effective than reactive approaches \cite{mitchell2019model, rakova2021responsible}. When issues arise in already-deployed systems, organizations typically need to fix the problem without disrupting service quality or availability. They might need to roll back to a previous version of the system that is less performant or apply quick patches that may not fully resolve the issue. Documentation that helps identify potential problems before deployment can lead to more comprehensive solutions than addressing issues post hoc once they are discovered in production systems \cite{holland2018dataset, ahlawat2024minimum}. 


Furthermore, documentation could preserve institutional knowledge. In any organization, particularly complex ones, consistently institutionalizing values and best practices can be challenging. Documentation could serve as a means of conveying technical information about systems, and the organization's policies and values that influence development and use \cite{ahlawat2024minimum}. For instance, documentation can aid in onboarding new employees by providing information about the systems they'll work on and communicating the organization's approach to development and governance \cite{adkins2022prescriptive, chang2022understanding, heger2022understanding}. 

Lastly, documentation could facilitate both internal and external audits \cite{brajovic2023model, crisan2022interactive, hupont2022documenting, miceli2021documenting, ahlawat2024minimum}. Audits may focus on the components of AI systems, the overall application, or the processes used in development \cite{mokander2023auditing}. Documentation could help organizations demonstrate that their claims—such as providing equitable system performance across different demographic groups—are backed by evidence. In these cases, documentation could guide auditors in selecting appropriate methods based on the auditor's objectives and the available data \cite{ahlawat2024minimum}. When audits examine the robustness of an organization's governance, documentation becomes even more critical since such information cannot be produced by analyzing the system itself \cite{clavell2024checklist}. Without comprehensive documentation, organizations may struggle to demonstrate the integrity and consistency of their processes.

\subsection{Challenges} 

Both proposed frameworks and empirical studies delineated how documentation could support more robust governance of AI systems; however, empirical studies, in particular, illuminated challenges organizations and practitioners often face, which could prevent the theories of change from being fully realized. These challenges included limited incentives and resources for robust documentation, structural and organizational barriers to communication, interpersonal and organizational constraints to ethical action, and poor integration with existing workflows, practices, and organizational infrastructure.

\subsubsection{Challenge \#1: Poor incentivization and limited resources can undermine documentation quality.} Practitioners acknowledge that the benefits of documentation could outweigh the costs of creating and maintaining it \cite{heger2022understanding}; however, when organizations fail to provide clear incentives for high-quality documentation, practitioners are less likely to prioritize it. This lack of motivation can hinder the realization of documentation’s potential benefits. 

Organizations typically do not prioritize documentation unless regulations require it for compliance or clients specifically request it \cite{chang2022understanding, miceli2021documenting, heger2022understanding}. Practitioners often view producing high-quality documentation as less relevant to their evaluations and promotions than development tasks that directly contribute to product outputs \cite{chang2022understanding, heger2022understanding}. This disconnect between the benefits of documentation and practitioner objectives is further exacerbated when the practitioners who benefit most from documentation (i.e., consumers) are not the same as those responsible for producing it (i.e., producers). In organizations where governance processes mandate documentation, a lack of familiarity with AI documentation concepts among developers could potentially magnify tensions between product development and governance teams \cite{ali2023walking, ahlawat2024minimum}. Moreover, practitioners are often more motivated by compliance than by normative considerations, which could lead to blind spots when novel ethical issues arise outside existing policies \cite{heger2022understanding}.

Under time constraints and competing priorities, practitioners may rush the documentation process, sometimes cutting corners to save time \cite{chang2022understanding, heger2022understanding}. For instance, they may reuse content from existing documentation, even when it describes different systems, resulting in incomplete or inaccurate records for the current system \cite{pushkarna2022data, heger2022understanding,{ahlawat2024minimum}}. Practitioners also sometimes leave questions unanswered rather than seek out the necessary information \cite{heger2022understanding} or prioritize recording information that is most valuable to them, potentially omitting critical details for others less familiar with the system \cite{hind2019experiences} or leaving downstream practitioners with an incomplete end-to-end understanding of systems \cite{ahlawat2024minimum}. Furthermore, because practitioners often forget relevant information if they do not record it during development, their reconstruction of key system information after the fact can be time-consuming and error prone \cite{reid2023right}. 

A potential consequence of rushed, low-quality documentation is that it may inadequately inform—or even misinform—downstream practitioners, potentially compromising effective development and risk management \cite{pepe2024huggingface, liang2024whats, geiger2020garbage}. For example, many publicly available model cards lack details about out-of-scope uses, limitations, or environmental impacts \cite{pepe2024huggingface, liang2024whats}. Less than 30\% of model cards contain evaluation results, a critical piece of information for downstream practitioners \cite{liang2024whats}. Also, some publicly available documentation contains incorrect license information, potentially leading practitioners to inadvertently violate license terms \cite{pepe2024huggingface}.

Practitioners frequently advocate for automating as much of the documentation process as possible to maintain its relevance and reduce its time demands \cite{rostamzadeh2022healthsheet, heger2022understanding, pushkarna2022data, ahlawat2024minimum}. However, certain essential details—such as the rationale behind specific design or development decisions—are challenging to automate. This type of information is often crucial for evaluating potential harms and for enabling accountability \cite{yurrita2023generating, sambasivan2021everyone}. Recognizing this, some practitioners acknowledged that parts of the documentation process must remain manual \cite{pushkarna2022data}. However, if these manually updated sections are not regularly synchronized with system updates, outdated information could itself become a source of harm \cite{mcmillan2021reusable}. As a result, deprioritizing documentation could lead to a situation where the most critical elements for risk management are the most likely to become outdated.



\subsubsection{Challenge \#2: Structural and organizational barriers can complicate collaboration} Several practical barriers can limit how well documentation can act as a springboard for active communication and collaboration.

The modern AI supply chain, in which general-purpose models are adapted to a variety of downstream tasks \cite{vaswani2017attention}, introduces additional structural complications for documentation. In some cases, different teams within the same organization may adapt a general-purpose model for specific tasks. In these instances, documentation might prompt the downstream team to directly reach out to the original development team to discuss the system’s capabilities and limitations or clarity details. However, in other cases, a team deploying a model may rely on a model developed by a different organization \cite{jones2024foundation}. In this case, it will typically be more difficult if not impossible for the downstream deployment team to directly communicate with the original development team. Any information that is missing or confusing in the documentation artifact cannot be easily clarified through collaboration.

Furthermore, when AI development implicates complex supply chain dynamics, actors further removed from the deployment context often struggle to account for the needs and considerations of external stakeholders \cite{balayn2024understanding}. This disconnect may manifest in risk documentation for upstream models that fails to accurately or comprehensively capture the real-world risks these systems pose in specific deployment contexts.

Challenges relevant to documentation of general-purpose models may be especially pronounced for less technical downstream stakeholders, who may struggle to understand the functionality and risks associated with these models \cite{crisan2022interactive}. More interactive forms of documentation, such as those that allow users to engage with the models directly or customize the level of detail presented, could help mitigate these issues by making the information more accessible and relevant \cite{crisan2022interactive, liao2023designerly, moore2023failure}, especially when direct communication between teams is limited. However, further research is needed to determine the most effective ways to document general-purpose systems while avoiding overwhelming documentation consumers with excessive or irrelevant details.

Beyond the structural challenges of general-purpose models, a major organizational challenge is that documentation that is misaligned with the needs of its intended consumers can complicate rather than facilitate stakeholder communication. If documentation artifacts are overly technical or jargony, documentation consumers who are less technical or less familiar with the system may struggle to understand what details are most relevant, or misinterpret described practices \cite{crisan2022interactive, liao2023designerly, miceli2021documenting, shen2022model}; if documentation lacks relevant details, documentation consumers may overlook risks that become apparent only later in the development process, which may require teams to substantially backtrack. In either case, mismatch in comprehension can contribute to needless disruption to normal workflows. 


\subsubsection{Challenge \#3: Interpersonal and organizational constraints can limit ethical decision making} In some cases, documentation can improve critical reflection \cite{boyd2021datasheets}; however, several practical barriers may limit the extent to which documentation can improve ethical deliberation and decision making. First, the assumption that documentation artifacts and processes can promote ethical reasoning among practitioners hinges on the belief that practitioners are sufficiently aware of AI's ethical risks to users, non-users, and society that they can recognize them when prompted. However, without proper training in responsible AI practices and exposure to groups potentially harmed by AI systems, practitioners may not connect documentation with ethical risks \cite{heger2022understanding}. Consequently, their consideration of harms may only partially encompass the scope of risks, especially those requiring non-technical mitigations. For example, practitioners often misunderstand how bias can manifest in their work, leading them to incorrectly state in their documentation that bias concerns are not relevant \cite{hind2019experiences}. While it may not be necessary for ethicists to be embedded in the work process for practitioners to make responsible choices, as some have suggested \cite{mclennan2022embedded}, findings point to the need for  more support and training for practitioners in how to identify, document, and effectively communicate potential risks \cite{heger2022understanding, madaio2024learning}.

Second, practitioners and organizations may resist documenting ethical impacts they \textit{have} identified. Publicly available documentation artifacts, such as Hugging Face model cards and GitHub repository README files, often show little consideration of ethical concerns \cite{pepe2024huggingface, liang2024whats}. Sometimes practitioners resist addressing ethical impacts because they feel unqualified to speculate on numerous potential use cases and their impacts \cite{crisan2022interactive, heger2022understanding, hind2019experiences, nunes2022using, ahlawat2024minimum}. Others worry that detailing ethical concerns might give downstream stakeholders a false sense of security \cite{crisan2022interactive}. This hesitation could also stem from a desire to avoid personal responsibility for their actions \cite{nunes2022using}. Alternatively, practitioners could plausibly have concerns that documenting potential ethical harms could create a legal or public relations risk if outside stakeholders gained access to documentation.

Third, even if practitioners recognize and record ethical concerns, they may have little influence over their organization's development goals, deliverables, and timelines \cite{ahlawat2024minimum}. As a result, they might be unable to make the changes to data, models, or systems that they have identified as useful or necessary to address ethical risks through the documentation process \cite{ahlawat2024minimum, miceli2021documenting}. One ethnographic study found that business demands, not the beliefs of data subjects or practitioners, largely determined the organization’s documentation approaches \cite{miceli2021documenting}. For example, although practitioners who reflected on data labeling practices recognized that socially constructed identities are complex, the organization nevertheless chose to represent identity in a reductive way, such as by defining race according to discrete, mutually exclusive categories. The authors concluded that explicit and implicit power structures among internal organizational stakeholders significantly affect practitioners' documentation practices and ability to shape outcomes. Documentation approaches must be responsive to these constraints. Otherwise, documentation about ethical considerations may not promote meaningful action or may be incomplete or misleading.

\subsubsection{Challenge \#4: Poor integration with technical and organizational infrastructure can undermine diffuse benefits}
Practitioners studies emphasize that documentation is likely to be more effective when it is well integrated with practitioner workflows \cite{heger2022understanding, bhat2023aspirations}. For example, when practitioners are asked to produce documentation using external tools, they are more likely to make errors and are less likely to make note of relevant ethical considerations than when documentation methods are well integrated with their existing tools \cite{bhat2023aspirations}.  

Beyond undermining the quality of documentation artifacts themselves, poor technical and organizational integration of documentation processes with other organizational infrastructure could limit the extent to which documentation serves as a forcing function for best practices in development and governance generally. In cases where documentation is ad-hoc and relies on tools that are not well-integrated with practitioners’ workflows \cite{bhat2023aspirations, heger2022understanding}, connecting documentation processes and artifacts to other aspects of the development and governance process becomes even more challenging {\cite{ahlawat2024minimum}. Without a solid documentation infrastructure that aligns with other organizational functions, documentation may fail to effectively enhance development and risk management.

Another challenge is that while high-quality documentation can have compounding benefits to governance and development, poor-quality documentation can have compounding negative impacts. Documentation should prevent redundant efforts, maximize time spent managing unique risks, and support proactive system development. However, these benefits depend on the quality of the documentation. Inadequate or incorrect documentation can lead to wasted time, overlooked risks, and unexpected issues that may require correction later in development. This could result in harms remaining unaddressed or inefficiencies undermining the overall governance process or quality and reliability of development.

\subsection{Design considerations}
How organizations choose to document AI systems reflects both their normative values and practical constraints. Design and implementation choices are also influenced by established and emerging external requirements, such as regulations that companies must adhere to. As organizations (and regulators) aim to maximize the benefits of documentation across various types of AI systems and components, they must often navigate tensions in their approach. Our analysis identified common design tensions or considerations organizations must navigate. These include: balancing customization and standardization, tailoring documentation to specific audiences, determining the appropriate level of detail, deciding on the degree of automation in the documentation process, and incorporating interactivity into documentation artifacts. To an extent, these design considerations are interrelated: the level of detail documentation artifacts contain can also be a form of audience tailoring, for example. And, audience tailoring can be viewed as a form of customization. Yet each of these design tensions at least partially distinct motivations and implications, so we address them separately.

\subsubsection{Design consideration \#1: Determining the degree of customization versus standardization in documentation artifacts.}
Organizations often create datasets, models, and systems for various purposes, employing a wide range of techniques and components. This diversity leads to a variety of risks, which customization is well-suited to address. At the same time, standardized documentation helps establish consistent practices and institutionalize norms. 

One advantage of customization is that it can address the specific risks and capabilities of an organization or AI system. For instance, annotating the origin of data collected in different healthcare settings can be crucial, as different settings serve distinct patient populations and employ varied healthcare practitioners. This information can help practitioners identify gaps in the dataset or regions where the dataset may be less applicable \cite{rostamzadeh2022healthsheet}. Conversely, annotating the origin of code snippets from software engineers working in different contexts may be less important since code typically functions similarly across different environments.

Echoing calls to ground approaches to AI harms within the context of deployment \cite{hutchinson2022evaluation, narayanan2024safety, nicholas2024grounding}, several frameworks recommend collectively documenting the system components that pertain to a given use case rather than separately documenting datasets, models, or systems independent of their intended use case \cite{chmielinski2020dataset, hupont2022documenting, mohammad2022ethics, rostamzadeh2022healthsheet, srinivasan2021artsheets}. For example, some researchers have proposed specific documentation methods for affective computing \cite{hupont2022documenting}, since these methods pose unique risks related to psychological manipulation and surveillance and so may merit a tailored approach. Moreover, because affective computing applications are likely to be classified as high-risk by the EU AI Act, applications using affective computing are also more likely to be subject to specific documentation requirements \cite{hupont2022documenting}. 

Customization can also be valuable at the model level, as different models are designed to perform unique functions and thus pose unique risks. In one study, practitioners argued that the information within a model card should be rearranged to ensure that documentation consumers clearly understand the specific model's purpose and limitations \cite{crisan2022interactive}. Another study found that practitioners added technical details to standardized documentation, even when instructed not to, suggesting that they believe standardized formats need flexibility to include model-specific details they find relevant \cite{bhat2023aspirations}.

At the organizational level, customization might also be necessary. For example, a healthcare company may require a different documentation approach than a financial company, and the approach either should take may depend on the company's maturity level \cite{dotan2024evolving}. In some cases, it may be appropriate to customize documentation at multiple levels to respond to both the organization’s nature and the type of system it is developing \cite{chmielinski2020dataset, gebru2021datasheets, holland2018dataset, richards2020methodology, roman2023open, stoyanovich2019nutritional}.

While customization has many benefits, it also presents challenges. Standard formats facilitate the development of tools that allow practitioners to search for system components that meet particular specifications, improving discoverability and efficiency within an organization’s ecosystem \cite{heger2022understanding}. Since such tools might enable practitioners to more easily find appropriate components, it might also reduce the likelihood that they employ inappropriate ones. This enhanced discoverability is often cited as one of the most significant benefits of documentation by practitioners \cite{heger2022understanding, reid2023right}.

Standardized documentation could also facilitate practitioners comparing candidate datasets, models, and systems, and for comparing AI and non-AI approaches \cite{chmielinski2024clear, gilbert2023reward, hind2019experiences, mitchell2019model, reid2023right, shen2022model, stoyanovich2019nutritional, zheng2022network}. When documentation artifacts vary significantly in the type of information they contain or how this information is presented, comparisons become more challenging. Given the time constraints and the currently limited incentives for creating \cite{chang2022understanding, heger2022understanding, miceli2021documenting} and using \cite{crisan2022interactive} documentation, maintaining the core utility of documentation through standardization could support broader adoption within organizations.

A more diffuse but critical benefit of standardization is its role in helping the AI community converge on norms of practice and communication \cite{arnold2019factsheets, mcmillan2021reusable, mcmillan2024data}. Practitioners often struggle with how to complete documentation or what level of detail to include \cite{chang2022understanding, pushkarna2022data, heger2022understanding, miceli2021documenting}, leading them to rely heavily on existing examples, which may not always be appropriate. Standardization can mitigate these issues by providing clear expectations for what information should be included and what consumers should expect to find in documentation artifacts, thus encouraging more consistent practices \cite{mcmillan2021reusable, pushkarna2022data, roman2023open}.

Standardization also might facilitate structured risk management. If documentation producers are instructed to select certain details from structured categories, or certain fields are constrained to particular structured formats, these fields can be used to trigger certain governance actions, such as scheduling a review or requiring that a certain mitigation be applied prior to proceeding. On the other hand, such process automation can mean that concepts and risks are oversimplified, and that processes or mitigations are recommended in cases where they may not be appropriate, while leaving other relevant risks to be overlooked.

The choice between customization and standardization is not always a strict binary but a spectrum, allowing for a combination of both. For example, organizations can create standardized templates that are adapted to different contexts, such as healthcare models versus those for software engineers, ensuring consistency while meeting specific needs {\cite{bhat2023aspirations, heger2022understanding, hind2019experiences, holland2018dataset, mcmillan2021reusable, pushkarna2022data, richards2020methodology}. However, organizations may face trade-offs when deciding on the level of standardization to employ in documentation. 

\subsubsection{Design consideration \#2: Determining how much to tailor documentation artifacts to specific audiences} Organizations could tailor their documentation to fit specific application contexts or systems. In addition, tailoring documentation to meet the needs of its intended audience can help ensure it achieves the desired governance outcomes. When designing documentation strategies, organizations must consider the diverse needs of multiple stakeholder groups \cite{micheli2023landscape}. Different stakeholders often require distinct forms of documentation to be effective. To meet these varied needs, organizations might choose to produce tailored documentation for each stakeholder group. Alternatively, organizations might opt for a single, general-purpose artifact that all stakeholders can use, though this might not be adequately tailored to a specific purpose.

A single-format artifact may not adequately serve any particular stakeholder group. For instance, documentation designed for data scientists might not be accessible or useful to UX professionals or other non-technical stakeholders \cite{liao2023designerly}. Similarly, documentation aimed at public transparency might lack the depth needed by internal decision-makers handling risk management. This highlights a key trade-off: while a single-format document can reduce the complexity of documentation management, it risks failing to meet the specialized needs of different groups.

Stakeholder-specific documentation, meanwhile, can address the unique needs of different groups of practitioners. These groups, including users, policymakers, data scientists, engineers, lawyers, and others \cite{micheli2023landscape}, often have distinct goals and use different terminologies, which can influence the type of documentation produced and the most helpful forms for each group. Documentation tailored to its specific audience is more likely to provide stakeholders with the necessary information for accomplishing their goals. Without a clear target audience, documentation producers might only address what is relevant to their own team \cite{heger2022understanding}, potentially overlooking critical information needed by other teams \cite{hind2019experiences} or presenting it in an unusable format \cite{liao2023designerly}.

Tailoring documentation is particularly important when conveying technical information to non-technical stakeholders, both within and outside the organization. While many researchers propose frameworks that aim to make documentation comprehensible to both technical and non-technical practitioners \cite{holland2018dataset, mcmillan2021reusable, mohammad2022ethics, pushkarna2022data, richards2020methodology, roman2023open, sokol2020explainability, tagliabue2021dag}, empirical studies indicate that non-technical audiences often struggle to understand even simplified technical details \cite{crisan2022interactive, liao2023designerly, miceli2021documenting, shen2022model}. For example, in a study where non-technical practitioners used documentation to assess model risks, a significant proportion failed to correctly identify the model’s basic purpose \cite{crisan2022interactive}. Using documentation, non-technical practitioners may struggle to grasp technical information, such as accuracy metrics \cite{shen2022model}. This suggests that unless technical information is presented in simpler terms, non-technical stakeholders may still face challenges in interpreting technical information necessary for risk assessment when it is not specifically tailored to their needs. 

Another advantage of stakeholder-specific documentation is its ability to leverage the expertise of different practitioners more effectively. Tailoring documentation allows stakeholders to focus on the most relevant information for their roles, thereby enhancing the organization’s ability to identify and manage risks. For instance, in developing a documentation framework for healthcare data, machine learning experts prioritized the dataset’s composition, while healthcare experts focused on the details of medical diagnoses and data collection processes, as these were crucial for interpreting the data’s relevance and limitations \cite{rostamzadeh2022healthsheet}. By emphasizing information pertinent to their expertise, stakeholders can conduct more thorough analyses and potentially uncover issues that others might miss \cite{pushkarna2022data, rostamzadeh2022healthsheet}.

Tailored documentation can also be more actionable for practitioners. One common issue with general-purpose documentation is that it may not clearly guide stakeholders on how to use the information provided \cite{adkins2022prescriptive, crisan2022interactive, stoyanovich2019nutritional}. This issue is particularly significant when practitioners fail to connect documentation with the ethical impacts of their work \cite{heger2022understanding}, indicating a need for prescriptive guidance to manage AI risks effectively. The actions practitioners take in response to documentation often depend on their specific roles for two reasons. First, when documentation is aligned with the purpose of their role, practitioners are better able to understand the necessary actions to manage risk. For example, information about how well a system meets user needs may be more actionable for a UX professional than a machine learning engineer \cite{liang2024whats}. Second, practitioners typically have the authority to act in specific ways within organizations. For instance, compliance teams cannot implement changes to the production codebase, and data scientists are not responsible for conducting legal reviews. Tailored documentation that aligns with each group's roles and responsibilities can guide them toward actions they are empowered to take. 

Whereas multiple forms of documentation can better meet diverse stakeholder needs, single-format documentation can help reduce confusion and avoid the fragmentation that occurs when information is scattered across different formats, such as README files, wikis, and slide decks \cite{heger2022understanding}. Even with centralized repositories that link different versions, managing multiple formats can still lead to confusion about the existence and authority of these artifacts and make it difficult for practitioners to access the information they need and develop a cohesive understanding.

Furthermore, single-format documentation may be easier to produce and maintain. If practitioners are required to create multiple forms of documentation to meet the needs of different stakeholders, organizations must ensure sufficient time is allocated for this process. Without adequate time, practitioners may rush, leading to low-quality documentation. In cases where time constraints are a concern, it may be more effective to focus on producing a single, high-quality document that is regularly updated. At a minimum, it may be beneficial if documentation artifacts clearly indicate their last update to help practitioners gauge relevance and accuracy.

\subsubsection{Design trade-off \#3: Determining the level of interactivity documentation should support}
Producing multiple forms of documentation is one way to support a variety of goals, stakeholders, and systems; however, these multiple forms of documentation create challenges for organizations to manage. Interactive documentation offers an alternative solution by enabling stakeholders to access the specific information they need while maintaining overall usability and comprehension. 

Interactive documentation may be able to effectively serve multiple stakeholders with a single artifact. For example, interactive system diagrams might allow practitioners to click on specific components to access technical details \cite{procope2022system}, while expandable sections can provide additional context as necessary \cite{crisan2022interactive}. 

Empirical research supports the idea that interactive documentation can significantly improve understanding, particularly among non-technical audiences \cite{crisan2022interactive, liao2023designerly, moore2023failure}. Tools like Hugging Face's model inference API\footnote{https://huggingface.co/docs/api-inference/} or OpenAI’s developer playground\footnote{https://platform.openai.com/playground} allow practitioners to engage with models and systems directly, fostering an intuitive understanding of how inputs and outputs are connected. This hands-on interaction is crucial for designing products that leverage the system’s capabilities and for assessing potential risks \cite{liao2023designerly}. Interactive access is particularly valuable when exploring pre-trained, general-purpose models, where functionality can vary significantly depending on the implementation \cite{liao2023designerly, moore2023failure, crisan2022interactive}.

However, interactive documentation also presents challenges. Without good information architecture, it can overwhelm users, leading to cognitive overload rather than enhancing understanding \cite{crisan2022interactive}. Additionally, interactive documentation requires additional engineering effort to design and maintain, which may be resource-intensive for organizations. In contrast, static documentation offers consistency and simplicity. It presents information in a uniform format, fostering a common understanding that can enhance communication, collaboration, and decision-making across teams.

\subsubsection{Design consideration \#4: Determining the level of detail documentation artifacts should contain.}
Organizations must determine what level of detail to include in documentation artifacts. Whereas exhaustive documentation might ensure practitioners have access to all the necessary information, concise documentation may improve ease of production and use.

Concise documentation could help stakeholders focus on the details that are most relevant to their needs, reducing the likelihood of information overload. For documentation aimed at downstream consumers, including internal risk management professionals, it is essential to provide enough detail to support decision-making without overwhelming them. Information overload can lead to selective attention, where decision-makers may focus on certain details while neglecting others, potentially degrading the quality of their decisions \cite{phillips2020decision}, particularly under time constraints \cite{hahn1992effects}. For example, faced with extensive and complex documentation on AI systems, consumers might not read documentation at all, miss critical information, or become frustrated and abandon the effort. As one participant in a research study noted, "I would lose patience after 30-40 seconds if I have to put a lot of effort into finding what I'm looking for” \cite{crisan2022interactive}. Concise documentation that highlights the most important information for a given stakeholder group can help internal decision-makers manage risk more effectively by ensuring they focus on critical details.

Moreover, concise documentation is likely easier to produce and maintain, which can enhance accuracy and encourage more frequent use. Exhaustive documentation may be burdensome to create and keep updated, particularly if the process is manual. For instance, in the initial case studies of an extensive data documentation framework, practitioners reported that completing the documentation took two to three hours, excluding the time needed to gather the required information \cite{bender2018data}. While a few hours might be manageable for a single project, this time commitment becomes overwhelming when scaled across an entire organization, potentially involving hundreds or thousands of datasets, models, or systems. In such scenarios, spreading effort thinly across all systems might detract from the focus on high-risk or high-impact components, undermining critical risk management efforts.

However, while concise documentation has its advantages, it may omit details that, although not relevant to most uses, are crucial for specific applications. Exhaustive documentation provides a more comprehensive overview that can accommodate a broader range of applications. For example, consider a dataset containing patient information and healthcare outcomes over a year. If a model uses this dataset to predict patients' health outcomes over time, the model's accuracy depends on the validity and reliability of the data's timestamps. Documenting how these timestamps were applied is essential for practitioners intending to use the dataset for this purpose. On the other hand, if the dataset is used for clustering patients into broad categories, temporal information becomes less critical. Therefore, the project's purpose and the specific stakeholders involved should shape the level of detail included in the documentation.

\subsubsection{Design consideration \#5: Determining the amount of automation in the documentation process.}
Documentation encompasses two main types of information: data that can be extracted automatically from the system and information that requires human input. Information that could potentially be derived directly from source code or generated through scripts interacting with system components includes elements such as data quantity, distribution, statistical properties, model types and parameters, system flow, and software library usage. In contrast, information related to human decision-making processes and organizational context requires manual input. This includes the motivations for developing specific datasets or models, criteria for selecting data sources, reasoning behind chosen methodologies, outcomes of compliance reviews, contact information for responsible parties, decommissioning procedures, and ethical considerations addressed during development. 

Providing detailed documentation of the rationale for design and development decisions underscores the subjective nature of these choices, which are often the underlying causes of AI-related harms \cite{yurrita2023generating, sambasivan2021everyone}. Manually created documentation can empower downstream stakeholders—and ultimately users—to scrutinize and contest the treatment they receive from AI systems  \cite{yurrita2023generating}. By transparently detailing how normative assumptions and subjective decisions are incorporated into system design and development, organizations can enhance accountability, supporting both internal governance and external stakeholder oversight.

One of the purposes of both producing and using documentation within an organization is to prompt developers to consider the ethical implications of their systems. However, the degree of automation in the documentation process can significantly influence how thoroughly practitioners engage with these considerations. Fully automated documentation requires minimal direct involvement, which may not encourage the level of critical reflection necessary to address complex ethical issues. As a result, some experts advocate for completing documentation manually, even when automation is feasible, to ensure that practitioners engage deeply with the material \cite{gebru2021datasheets, mcmillan2024data}.

However, automated documentation offers significant benefits in terms of efficiency and accuracy. Manually creating documentation is time-consuming and prone to errors, particularly if it is not well integrated with practitioners' existing tools and workflows \cite{bhat2023aspirations}. Practitioners have noted that automation can reduce the time required to complete documentation and increase the likelihood that it is kept up-to-date \cite{heger2022understanding}. Automation also reduces the burden on practitioners to recall critical information after the fact \cite{hind2019experiences}, which can be particularly valuable in complex, fast-paced environments. Organizations should also be aware that practitioners facing multiple competing priorities may circumvent manual processes by implementing their own forms of unaccountable automation. Well-designed automation of some portions of documentation could reduce this risk.

\section{Limitations}
Our research design and analysis have several limitations. First, we did not directly study the implementation of any specific documentation framework. Consequently, our assessment of the utility of our theories of change, and the challenges associated with them, is based on indirect analysis of existing work rather than direct evaluation.

Another limitation is the broad scope we applied to evaluating documentation. Different domains of documentation, such as documentation for voice data versus art data, documentation for specific system components, such as data versus models, and documentation for different stages of development, such as pretraining data versus model fine-tuning data, may have distinct impacts on governance. Moreover, as with other aspects of governance \cite{rakova2021responsible, winecoff2022artificial}, the role of documentation likely varies across industries and may function differently depending on an organization’s size and level of governance maturity \cite{ahlawat2024minimum, rakova2021responsible}. Our broad approach limits the depth of our analysis in addressing these specific nuances.

Lastly, our work did not explore how multiple forms of transparency could support or improve AI governance. Documentation is one of many tools that can provide transparency into AI systems and support internal and external accountability. In some cases, system- or component-level documentation alone may not sufficiently enable practitioners to understand how systems operate, the risks they pose, or why they might behave in unexpected or potentially harmful ways. In these cases, explainability methods or other forms of transparency may be necessary to provide the insight needed to understand risks or resolve outstanding issues.

\section{Directions for future research}
Given that organizations have unique requirements and operate in vastly different contexts, there is no one-size-fits-all approach to documentation or a universal set of information that applies to all datasets, models, or systems. Our analysis of proposed frameworks and empirical findings suggests several potential pathways organizations could explore to develop documentation practices tailored to their specific needs and constraints. Further design and empirical research in key areas could address the limitations of our study or expand into other promising directions. Such work could provide organizations across varying maturity levels and domains with more specific, actionable guidance for designing documentation processes that strengthen their AI governance.


\textbf{Leveraging empirical evidence and feedback for continuous improvement.} Our sample of publications contained significantly higher number of proposed documentation frameworks compared to empirical evaluations of their effectiveness, highlighting gaps in understanding how theoretical approaches to documentation can be translated into practical, impactful practices within specific organizational contexts. Notably, most empirical findings on documentation were cross-sectional, offering limited insight into how iterative evaluation and refinement could enhance documentation practices over time within AI organizations.

Future empirical research could fill this gap by identifying how organizations can evaluate potential documentation approaches via pilot projects and periodically collect empirical data on the efficacy of their documentation practices in meeting the needs of relevant stakeholders. Research could also inform how organizations can plan to assess the usability of documentation frameworks and their effectiveness in achieving the organization's goals \cite{berman2024scoping}, and adapt their documentation strategies by conducting continuous assessments as their needs or technology evolves. Beyond scholarly research efforts, given the limited empirical data on the effectiveness of documentation approaches, organizations can also contribute to the broader AI community by publicly sharing their findings on what works best in different contexts. Organizations may benefit from empirical evidence and key performance indicators that are both qualitative and quantitative. Quantitative metrics are often best suited to capturing broad patterns across many instances but may lack depth for individual phenomena. Qualitative measures provide "thick" insights for specific instances \cite{geertz2008thick} but may have limited generalizability.

\textbf{Identifying unexpected negative consequences of design choices.} Empirical findings from researchers or organizations on how theoretically sound design choices—including those discussed in this work—might lead to unintended negative consequences can offer valuable insights for shaping best practices grounded in real-world experiences. For instance, a study on practitioners might reveal that while interactive interfaces help non-technical users understand information, they could hinder technical practitioners from quickly accessing relevant details. Similarly, research on practitioners or publicly available artifacts might show that practitioners often omit critical information, even when documentation is designed to be concise. Such findings could inform revisions to the frameworks and design considerations discussed in this work.


\textbf{Informing the degree and design of standardization of artifacts}. Our research highlights the advantages of standardization including enabling comparison between documentation artifacts, facilitating search and discoverability of information, establishing consistent expectations for documentation producers and consumers, and institutionalizing norms of practice. For these reasons, some level of standardization in documentation processes and artifacts is likely to be helpful for most organizations for most AI systems. However, standardized formats may overlook or insufficiently address the unique characteristics of specific AI system components or categories (e.g., language versus vision models), or application domains (e.g., social media versus healthcare). Future research could benefit more specific applications by providing insight into the appropriate level of standardization and identifying when and to what extent deviations from standardized formats are advantageous.

One approach to balancing standardization and customization that research has begun to explore is having customizable documentation templates or interchangeable modules  \cite{bhat2023aspirations, heger2022understanding, hind2019experiences, holland2018dataset, mcmillan2021reusable, pushkarna2022data, richards2020methodology}. Templates could, for example, allow for customization of subsections while maintaining consistent content and structure within those subsections. For instance, companies may include a subsection on the labeling process for AI systems based on supervised learning but exclude it for unsupervised AI systems since they do not require labels. Future research could explore how templates or other customization approaches can best maintain the usability and institutional benefits of standardization, while allowing for enough customization to address unique risks. 


\textbf{Designing documentation for general-purpose AI systems.} Although initiatives such as the EU AI Act require general-purpose model providers to supply downstream deployers with certain technical documentation to support their risk management, research suggests that the properties of upstream or base models may not always be relevant to downstream deployments \cite{weidinger2023sociotechnical, solaiman2023evaluating, qi2023fine}. Although approaches for documentation general-purpose models are emerging \cite{bommasani2023foundation}, disconnects between upstream models and downstream applications pose significant challenges for designing effective documentation recommendations for governing general-purpose systems.

For example, when downstream practitioners cannot directly communicate with upstream model developers, interpreting the general documentation in the context of specific applications can be difficult. Several studies have shown that interfaces allowing practitioners to observe system outputs in response to given inputs can enhance their understanding of model capabilities and risks \cite{liao2023designerly, crisan2022interactive, moore2023failure}. Consequently, interactive interfaces can serve as a valuable complement to traditional, static documentation. 

Research that illuminates how traditional, static approaches to documentation can be combined with interactive interfaces and even explainability methods may shed light on how documentation of general-purpose systems can best be communicated throughout the AI development lifecycle. Such work may be especially helpful in designing information architectures for documentation for practitioners without deep AI expertise, since these practitioners often face challenges in understanding complex technical information in model documentation \cite{crisan2022interactive}.

\textbf{Disentangling the role of different types of documentation in AI governance.} Our analysis reveals that AI documentation can broadly support a range of governance objectives through four general mechanisms. However, different types of documentation—such as data, model, system, or process documentation—either individually or in specific combinations, may serve distinct functions. For example, data documentation provides practitioners with insights into how inputs might influence system behavior. Details about the dataset—such as who is represented and what information is contained within data instances—can alert downstream practitioners to potential biases or risks, including the possibility of unfair outcomes or privacy violations. 

Model documentation, on the other hand, can help practitioners evaluate the reliability of system performance or make informed decisions about usage, considering factors like environmental impacts, which can be significant even during model inference \cite{chien2023reducing}. System and process documentation plays a crucial role in providing operational clarity, enabling downstream stakeholders or organizational decision-makers to assess deployment contexts, overarching system behaviors, performance, risks, and the adequacy of ongoing monitoring processes.

To ensure maximum effectiveness, the form and content of documentation artifacts—as well as the processes involved in creating them—should be responsive to the unique impacts and governance challenges associated with different system components. Future research, particularly empirical studies focused on organizational contexts, could clarify how documenting different system components—individually or in combination—supports distinct or overlapping governance goals. Such analysis could also inform design recommendations for different types of documentation, ensuring they collectively maximize their overall impact on governance.


\textbf{Institutionalizing responsible AI and risk management practices beyond documentation.} 
Ethical deliberation is a skill that practitioners can develop over time with intentional practice \cite{vallor2018overview}, but organizations need to implement additional interventions beyond documentation to help cultivate this skill. Even when practitioners engage in documentation processes, they may overlook critical risks or fail to recommend appropriate mitigations without a deeper understanding of how AI can harm diverse individuals and communities \cite{madaio2024learning, chang2022understanding}. Therefore, organizations should strive to embed responsible AI and risk management efforts throughout their tools and AI development practices. When responsible AI is integrated into the organization's culture and broader practices, documentation processes that prompt practitioners to reflect on risks are more likely to be informed by a comprehensive understanding necessary for effective risk management.

An open research question is how different tools and methods for raising awareness of ethical issues, fostering ethical deliberation, and empowering practitioners to act on these insights can work together to enhance AI governance. For instance, fairness toolkits can help practitioners less-experienced in algorithmic fairness better recognize and deliberate on trade-offs between system fairness and other performance indicators \cite{balayn2023fairness}. However, the tools and processes organizations adopt to institutionalize risk management practices may inadvertently narrow practitioners’ focus, limiting their ability to address emergent problems that fall outside the scope of these tools and processes or encouraging them to apply risk management techniques in inappropriate contexts \cite{balayn2023fairness, ahlawat2024minimum}.

Future research could explore how documentation and other AI risk mitigation tools can enable practitioners to manage risks consistently and rigorously while remaining flexible enough to identify and address new and unforeseen concerns. Research on how explainability methods can support documentation may be especially helpful in this regard. Stakeholders seek explanations for a variety of reasons, such as translating system predictions into concrete actions, ensuring compliance with relevant policies and regulations, understanding the impact of different model factors on their outputs, among others \cite{suresh2021beyond}. While documentation could alert practitioners to possible issues that might arise in deployment, it does not necessarily enable causal understanding of issues. Explainability methods, in contrast, can support practitioners in understanding model behavior at the instance level, helping them debug models or systems \cite{bhatt2020explainable}. Explainability methods may be especially useful in for helping practitioners conceptualize the behavior of general-purpose AI systems such as large language models (LLMs) \cite{liao2023ai}. By identifying which stakeholders would benefit most from explanations and what their goals are, organizations can design explainability approaches that work in tandem with documentation to support responsible risk management and decision making.


\section{Conclusion}
This paper underscores the critical role of documentation in enhancing AI governance within organizations, emphasizing that effective documentation practices are essential for managing AI risks and fostering responsible system development. To support robust governance, organizations must tailor their documentation processes and artifacts to meet the specific needs and constraints of their stakeholders. Establishing clear success criteria—such as accuracy, comprehensiveness, and usability—and regularly assessing progress against these goals is crucial for maintaining the effectiveness of documentation strategies.

While our focus has been on documentation for internal governance, these processes and artifacts can also form the foundation for transparency efforts directed at external stakeholders. By cultivating strong internal documentation practices and continually evaluating their success, organizations can build the necessary infrastructure to create detailed, actionable records of system development and risk management. These records are necessary for communicating effectively with external audiences, such as regulators and the public.

However, documentation designed for internal use may not seamlessly translate to external contexts. The differences in expertise, needs, and objectives between internal and external stakeholders require careful consideration. Organizations must adapt their documentation processes and artifacts to bridge these gaps, ensuring that they are both accessible and informative to all relevant parties. By doing so, organizations can enhance their governance efforts and contribute to a more transparent and accountable AI ecosystem.

\begin{acks}
We would like to thank members of the AI Governance Lab team at CDT and Emily McReynolds for their valuable feedback on our research design, analysis, and synthesis. We also extend our thanks to Samir Jain and Drew Courtney for their insightful comments on our manuscript. Finally, we are grateful to the participants of the "Improving Documentation for AI Governance" workshop, held at CDT in June 2024.
\end{acks}

\bibliographystyle{ACM-Reference-Format}
\bibliography{bibliography}


\end{document}